\pdfoutput=1
\documentclass{aa}  

\usepackage{graphicx}

\usepackage{caption}
\usepackage{subcaption}

\usepackage{txfonts}
\begin{document} 

   \title{Dissecting star-formation in the "Atoms-for-Peace" galaxy:}


\subtitle{UVIT observations of the post-merger galaxy NGC7252}


\author{K. George\inst{1}\fnmsep\thanks{koshy@iiap.res.in}, P. Joseph\inst{1,2}, P. C{\^o}t{\'e}\inst{3}, S. K. Ghosh\inst{4,5}, J. B. Hutchings \inst{3}, R. Mohan \inst{1}, J. Postma \inst{6}, K. Sankarasubramanian\inst{1,7}, P. Sreekumar \inst{1}, C. S. Stalin \inst{1}, A. Subramaniam\inst{1}, S.N. Tandon \inst{1,8}}


\institute{Indian Institute of Astrophysics, Koramangala II Block, Bangalore, India \and Department of Physics, Christ University, Bangalore, India \and National Research Council of Canada, Herzberg Astronomy and Astrophysics Research Centre, Victoria, Canada \and National Centre for Radio Astrophysics, Pune, India \and Tata Institute of Fundamental Research, Mumbai, India \and University of Calgary, Calgary, Alberta Canada \and  ISRO Satellite Centre, HAL Airport Road, Bangalore, India \and Inter-University Center for Astronomy and Astrophysics, Pune, India}


 
  \abstract
   {The tidal tails of post-merger galaxies exhibit ongoing star formation far from their disks. The study of such systems can be useful for our understanding of gas condensation in diverse environments.}
   {The ongoing star formation in the tidal tails of post-merger galaxies can be directly studied from ultraviolet (UV) imaging observations.}
   {The post merger galaxy NGC7252 ("Atoms-for-Peace" galaxy) is observed with the Astrosat UV imaging telescope (UVIT) in broadband NUV and FUV filters to isolate the star forming regions in the tidal tails and study the spatial variation in star formation rates.}
   {Based on ultraviolet imaging observations, we discuss star formation regions of ages $<$ 200 Myrs in the tidal tails. We measure star formation rates in these regions and in the main body of the galaxy. The integrated star formation rate of NGC7252 (i.e., that in the galaxy and tidal tails combined) without correcting for extinction is found to be 0.81 $\pm$ 0.01 M$_{\odot}$/yr. We show that the integrated star formation rate can change by an order of magnitude if the extinction correction used in star formation rates derived from other proxies are taken into consideration. The star formation rates in the associated tidal dwarf galaxies  (NGC7252E,  SFR=0.02 M$_{\odot}$/yr and NGC7252NW,  SFR=0.03 M$_{\odot}$/yr) are typical of dwarf galaxies in the local Universe. The spatial resolution of the UV images reveals a gradient in star formation within the tidal dwarf galaxy. The star formation rates show a dependence on the distance from the centre of the galaxy. This can be due to the different initial conditions responsible for the triggering of star formation in the gas reservoir that was expelled during the recent merger in NGC7252.}
  
  {}

\keywords{galaxies: star formation -- galaxies: interactions -- galaxies: dwarf -- galaxies: formation -- ultraviolet: galaxies}

\titlerunning{Dissecting the ongoing star-formation in "Atoms-for-Peace" galaxy}
\authorrunning{K. George\inst{1}}

\maketitle
%

\section{Introduction}

Galaxies in the local Universe are often classified morphologically as early-type (E/S0) and late-type galaxies (spirals). In a galaxy color-magnitude plot, the early-type galaxies fall on a `red sequence'
while late-type galaxies populate a `blue cloud'  \citep{Visvanathan_1977,Baldry_2004}. The red sequence galaxy population is observed to increase in numbers over cosmic time, as they evolve morphologically   \citep{Bell_2004,Faber_2007}. Galaxy merging is the most frequent morphological transformation process in low-density environments such as field and galaxy cluster outskirts: i.e., a merger of two gas-rich  spiral galaxies can result in the formation of an elliptical galaxy \citep{Toomre_1972}. The tidal forces involved can expel copious amounts of  molecular gas into the intergalactic medium, which later
condenses and triggers star formation \citep{Schweizer_1978}. Star formation in the tidal tails of merging galaxies is thus strong observational support for this hypothesis. The study of such star formation in the  intergalactic medium may be useful in understanding how star formation proceeds in low-density environments.\\

The stars in galaxies are normally formed from the dissipative collapse of baryons trapped within dark matter halos \citep{Blumenthal_1984}. The tidal tails of merging galaxies on the other hand are presumably devoid of dark matter halos, and hence represent a different environment for star formation. The mass and size of the gas condensations in tidal tails (particularly at their ends) are similar  to dwarf galaxies found in the local Universe, and are known as Tidal dwarf galaxies (TDGs) \citep{Zwicky_1956,Barnes_1992,Duc_1994,Duc_1998,Duc_2000,Duc_2007,Duc_2012,Kaviraj_2012}.  TDGs, although kinematically detached, are gravitationally bound to their parent galaxy and are observed to host large reservoirs of atomic and molecular gas \citep {Hibbard_1994,Braine_2000,Braine_2001}. The triggered star formation from the gas in the TDGs may thus help in understanding the formation of dwarf galaxies at high redshift \citep {Metz_2007,Kroupa_2010}.

NGC7252 ("Atoms-for-Peace" galaxy) is an archetypical advanced merger galaxy with a single-nucleus merger remnant, its tidal tails spread in almost opposite directions. NGC7252 is considered to  satisfy the "Toomre Sequence" of merging, where the merger remnant will eventually deplete the fuel for star formation and evolve into an elliptical galaxy \citep {Toomre_1972,Toomre_1977,Schweizer_1982}. Previous studies have demonstrated that NGC7252 is the remnant of an advanced merger between two gas-rich spiral galaxies, with two tidal tails and a main body of the galaxy, exhibiting shells and ripples in optical images\citep{Schweizer_1982,Dupraz_1990,Wang_1992,Fritze_1994,Hibbard_1994}. The merger is estimated to have started 600-700 million years ago and the associated starburst in the galaxy considered to be the result on that timescale \citep{Hibbard_1995,Chien_2010}. The surface brightness distribution of NGC7252 follows a roughly de  Vaucouleurs profile, with optical spectroscopy revealing post-starburst features \citep{Schweizer_1982,Hibbard_1999}. The galaxy also falls on the Faber-Jackson relation and has properties  similar to  elliptical galaxies located on the Fundamental Plane \citep{Lake_1986, Hibbard_1995,Genzel_2001,Rothberg_2006}.

The main body of NGC7252 also contains young star clusters triggered by the merger \citep{Miller_1997,Bastian_2013}. Star clusters formed in the tidal tails were studied extensively and the eastern and northwestern tidal tails were found to terminate in concentrations of neutral hydrogen that are coincident with the position of candidate TDGs \citep{Hibbard_1994,Knierman_2003}. The luminous blue star clusters  in the inner region of NGC7252 were found to have a narrow range in optical color ($V-I$ $\sim$ 0.65) with ages corresponding to 650–750 Myr, matching the numerical simulation of the time since the  start of the merger \citep{Knierman_2003}. The tidal tail of NGC7252 had been studied for its HI content, kinematics, H$\alpha$ kinematics, and gas phase metallicity \citep{Hibbard_1994,Lelli_2015}.  The neutral hydrogen observations of NGC7252 tidal tails reveal that the HI is spatially extended in width and length compared with the stellar populations in the optical image. The gas phase metallicity in  the tidal tail and the TDGs were found to be of solar value with no gradient indicative of star formation in the pre-enriched gas thrown out of the parent galaxy \citep{Lelli_2015}.

The ongoing star formation in the tidal tails of NGC7252 is thus of dwarf galaxy size (TDGs) and examining it at high spatial resolution is of interest for studying galaxy formation and evolution in diverse  environments.  The stellar populations in tidal tails can have old components retained from the pre-merger galaxies along with newly formed stars. The relative contribution (particularly of young stars) is difficult to disentangle from optical observations alone, but can be done with ultraviolet (UV) imaging. The integrated spectral energy distribution of young stellar populations peak at UV  wavelengths due to the presence of hot OBA stars. The UV flux directly traces the star formation within the past 100-200 Myrs \citep{Kennicutt_2012}.

The observations of NGC7252 presented here come from the NUV and FUV
channels of the ultraviolet imaging telescope (UVIT) on board ASTROSAT.  These have an angular resolution of 1\farcs2 in the NUV and 1\farcs4 in the FUV. The FUV flux represents the direct tracer of recent  and ongoing star formation. In a scenario of continuous star formation, it can thus constrain the star formation rate better than H$\alpha$ observations (which arise from star formation on timescales of $<$ 10 Myrs)\footnote{The presence of dust can however attenuate the emitted H$\alpha$ and FUV flux from the star forming regions.}.  We report here  deep UV imaging of NGC7252 that reveals star forming regions both in the tidal tail and in the main body of the galaxy, with good spatial resolution.

 The paper is arranged as follows. The UV observations and the analysis of tidal features, including an estimate of the star formation rate, is presented in \S2. We discuss the implications of our observations in  \S3 and  conclude in \S4. Throughout, we adopt a flat Universe cosmology with $H_{\rm{o}} = 71\,\mathrm{km\,s^{-1}\,Mpc^{-1}}$, $\Omega_{\rm{M}} = 0.27$, $\Omega_{\Lambda} = 0.73$ \citep {Komatsu_2011}.\\

\section{Observation, Data and Analysis}
The post-merger galaxy NGC7252\footnote{$\alpha$(J2000) = 22h20m44.7s and $\delta$(J2000) = 24d40m42s according to https://ned.ipac.caltech.edu/} has a spectroscopic redshift $\sim$ 0.0159, and is thus located at a luminosity distance\footnote {http://www.astro.ucla.edu/~wright/CosmoCalc.html} $\sim$ 68 Mpc \citep{Rothberg_2006}. The angular scale of 1" corresponds to 0.32 kpc at the distance of the galaxy.
NGC7252 was observed in FUV and NUV  wavelengths with UVIT on board the Indian multi-wavelength astronomy satellite ASTROSAT  \citep{Agrawal_2006}.
UVIT consists of twin telescopes, a FUV (130-180nm) telescope and a NUV (200-300nm)/VIS (320-550nm) telescope. The primary mirror is of 38cm diameter and generates circular images with a 28$'$ diameter simultaneously in all three channels \citep{Kumar_2012}.

There are narrow and broad band filters in each channel, of which we used the NUV N242W ($\lambda_{mean}$=241.8nm, $\delta\lambda$=50nm) and  FUV F148W ($\lambda_{mean}$=148.1nm, $\delta\lambda$=78.5nm)  for observations \citep{Annapurni_2016,Tandon_2017a}. The detectors of UVIT operate in photon counting mode, using a 512 $\times$ 512 CMOS detector. The  centroid positions are corrected for spacecraft drift and yield a resolution of $\sim$ 1\farcs2 for the NUV  and $\sim$ 1\farcs4 for the FUV channels.
The NUV and FUV images are flat  fielded  and corrected for distortion \citep{Girish_2017} using the software package CCDLAB \citep{Postma_2017}. The images from multiple orbits are co-added to create final images with total integration times: FUV =  8138~sec, and NUV = 7915~sec.The astrometric calibration is performed using the {\tt astrometry.net} package where solutions are performed using USNO-B catalog \citep{Lang_2010}. The  photometric calibration is  done using the zeropoint values generated from calibration stars \citep{Tandon_2017b}.

The optical $B-$band image of NGC7252 taken with CTIO 4-m telescope at a spatial  resolution of 1\farcs2 is used for comparing with UV images (data from \citet{Hibbard_1994}). We note that GALEX observations of NGC7252 exist, but with a relatively shallow depth (FUV net in
tegration  time = 560~sec, NUV net integration time = 1600~sec). Our UVIT imaging data for 
NGC7252 is deeper and at a higher spatial resolution. (GALEX resolution $\sim$ 4-5 arcsec.)\footnote{UVIT NUV N242W and FUV F148W filters have similar bandpass to GALEX NUV and FUV filters}
The foreground extinction from the Milky Way Galaxy in the direction of NGC7252 is $A_{V}$ = 0.10 \citep{Schlegel_1998} which we scale to the FUV and NUV
 $\lambda_{mean}$ values using the \citet{cardelli_1989} extinction law.

Star-forming regions may contain significant amounts of dust which can further attenuate the flux at UV  wavelengths. As we do  not have a measurement of the rest-frame extinction along the tidal tails and main body of NGC7252, we caution that our estimates for the star formation rates based on these FUV observations should be  treated  as lower limits to the actual values.


\subsection{Optical and UV analysis}

\begin{figure}
\centering
\includegraphics[width=8.0cm,height=12.0cm,keepaspectratio]{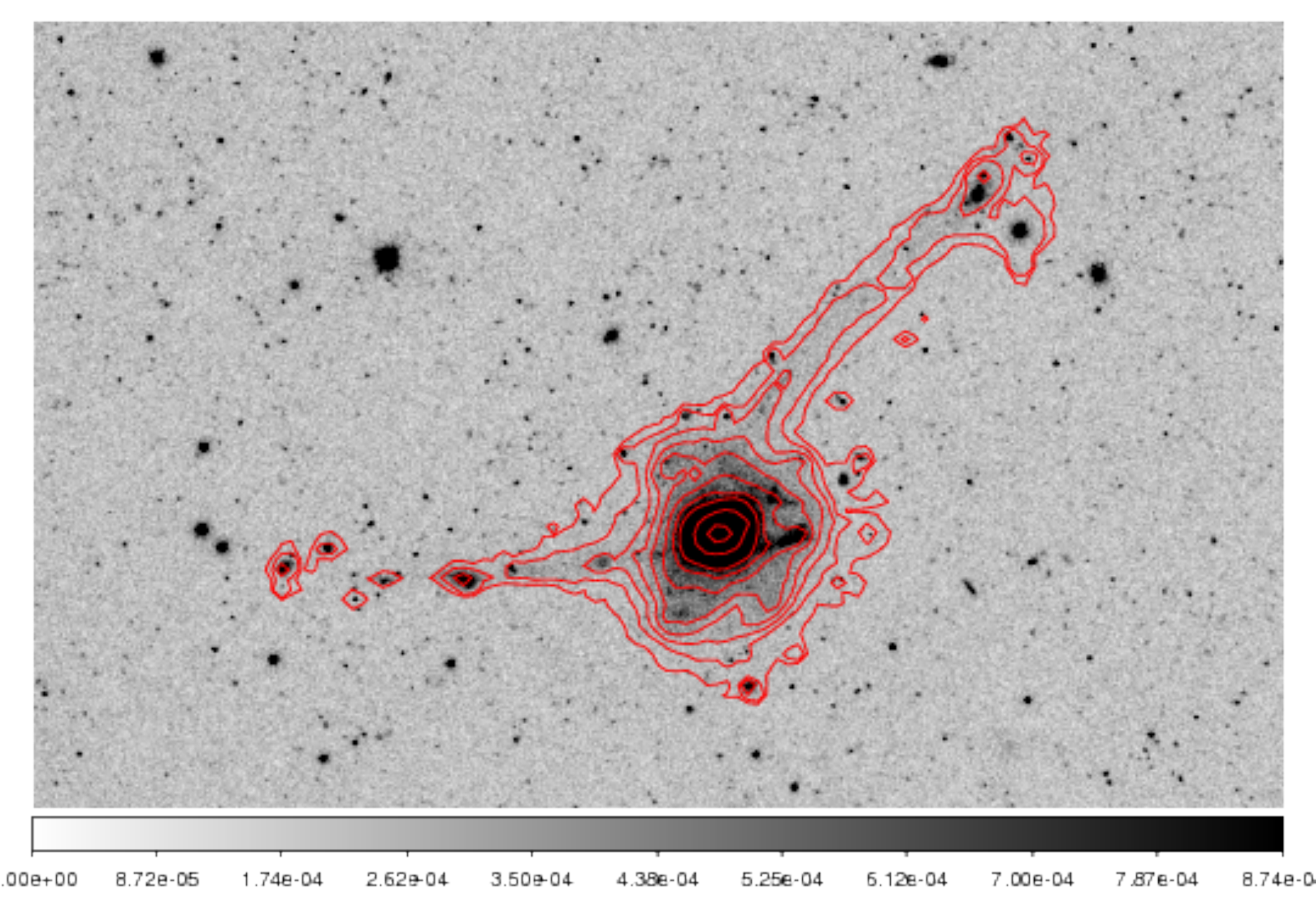}
\caption{The NUV image of NGC7252 with intensity values in inverted grey colour scale. The eastern and northwest tidal tails extending from the main body of the galaxy
 are clearly seen. The potential star  forming regions are clearly visible in the tidal tails. The optical $B-$band contour is overlaid over the NUV image in red. The NUV emission along the tidal tails is confined to narrow regions within
optical image. The image is of size $\sim$ 12.0$'$ $\times$ 8.0$'$ and the grey scale is in units of counts per sec.}\label{figure:fig1}
\end{figure}

The optical $B$-band image used in this analysis has a resolution that matches the NUV image, so it is possible to make a comparative study between the optical and NUV data.  The NUV  image is Gaussian-smoothed using a kernel of width three pixels (1\farcs2) to suppress noise. The NUV image of NGC7252 is shown in Figure~\ref{figure:fig1}.
The contours from the optical $B$-band image  are overlaid over the NUV image to highlight common features and identify possible missing features.  We note that all features are visible in both images along tidal tails and the main body of the galaxy. The underlying stellar  population in tidal tails can then be considered as young (< 1 Gyr) and due to the triggered star formation  since the onset of merger, as is evident from the NUV and optical imaging data.

\subsection{Star formation regions in the UV image}

\begin{figure*}
\centering
\includegraphics[width=15.0cm,height=15.0cm,keepaspectratio]{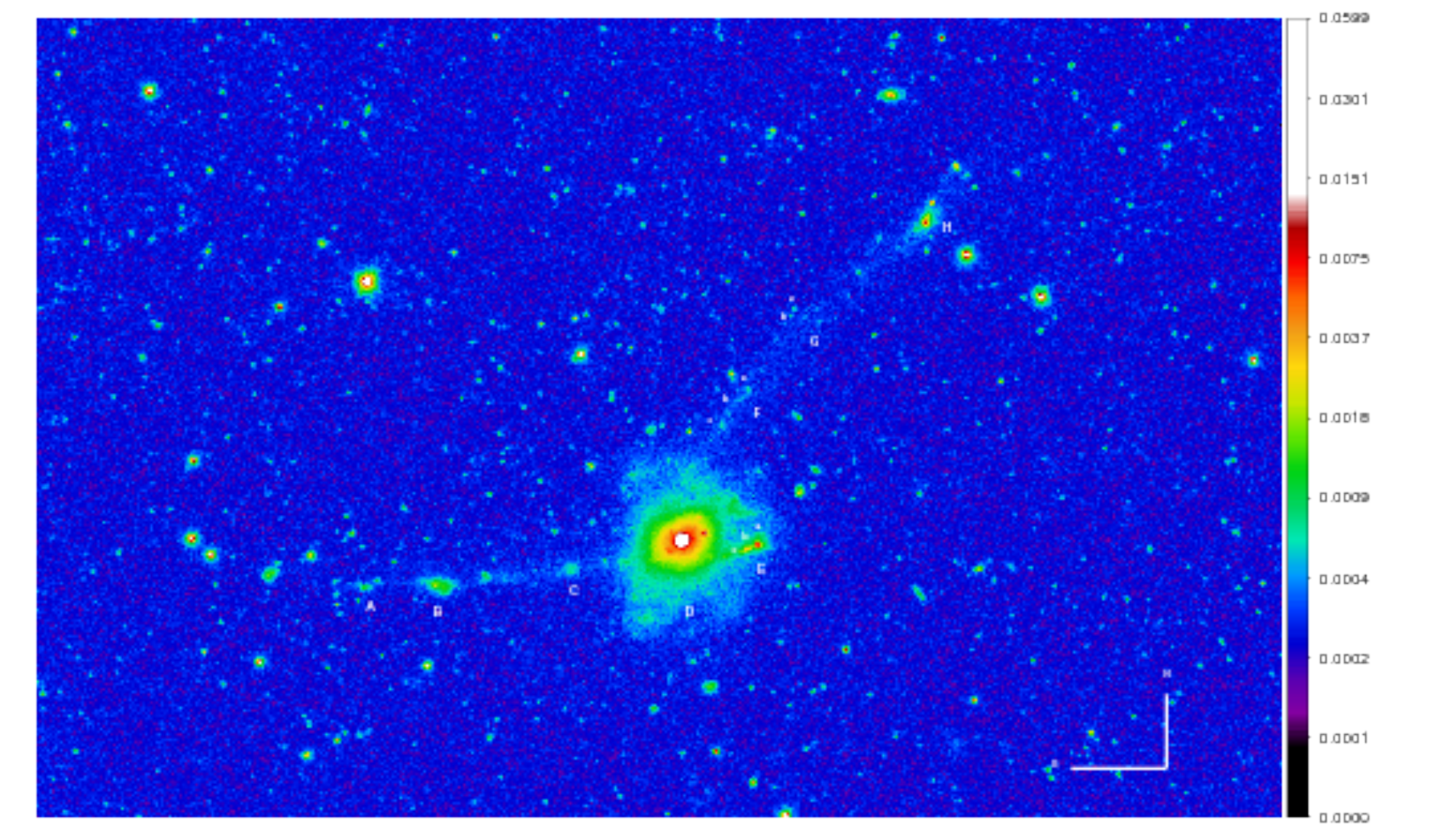}
\caption{The NUV image of NGC7252 with a contrast level chosen to highlight the low surface brightness level features in the tidal tails. The tidal features are marked while the colour scaling is shown in counts per second.}\label{figure:fig2}
\end{figure*}

\begin{figure*}
\renewcommand*\thesubfigure{\arabic{subfigure}}
\centering
\begin{subfigure}{0.3\textwidth}
  \centering
  \includegraphics[width=1.\linewidth]{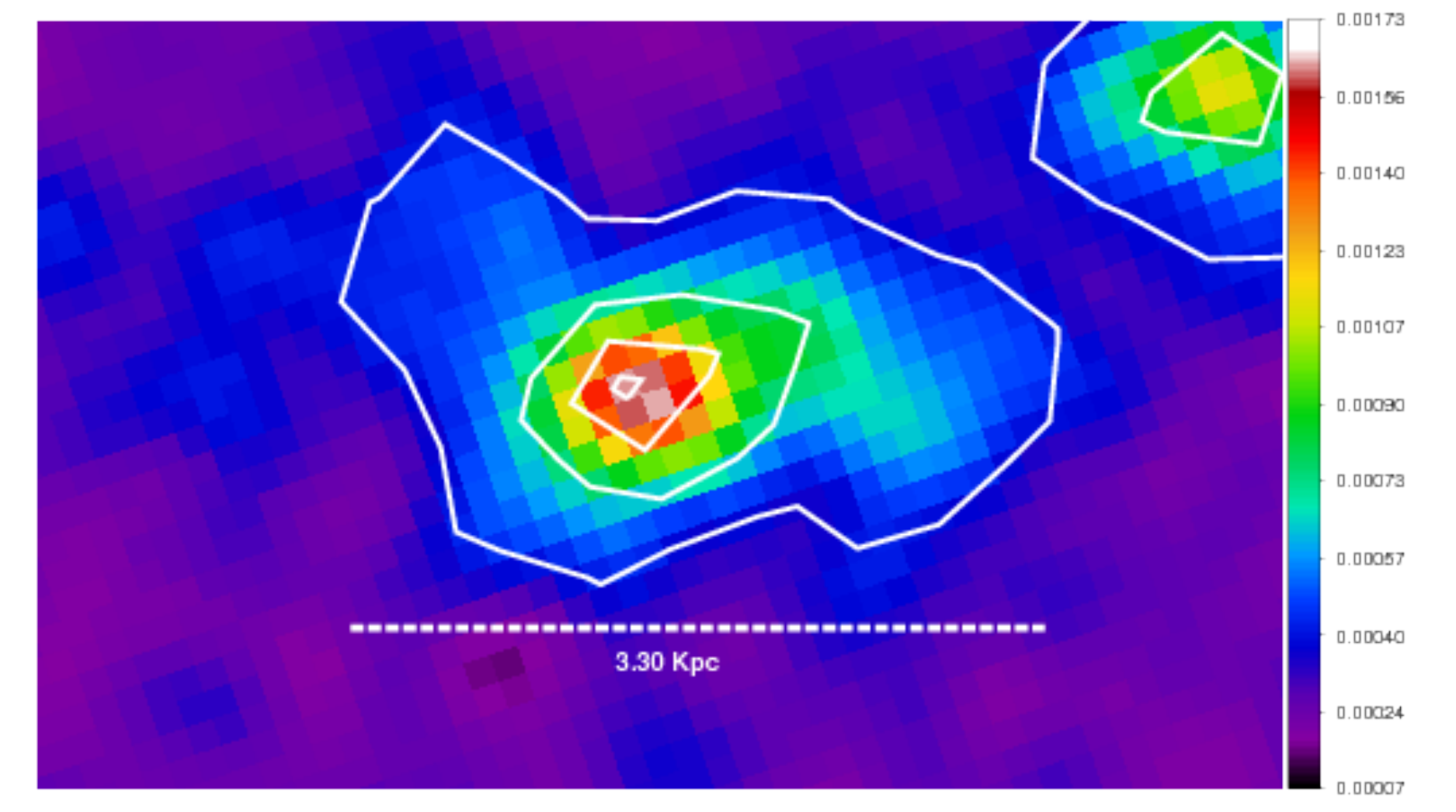}
  \caption{A}
\end{subfigure}%
\begin{subfigure}{0.3\textwidth}
  \centering
  \includegraphics[width=1.\linewidth]{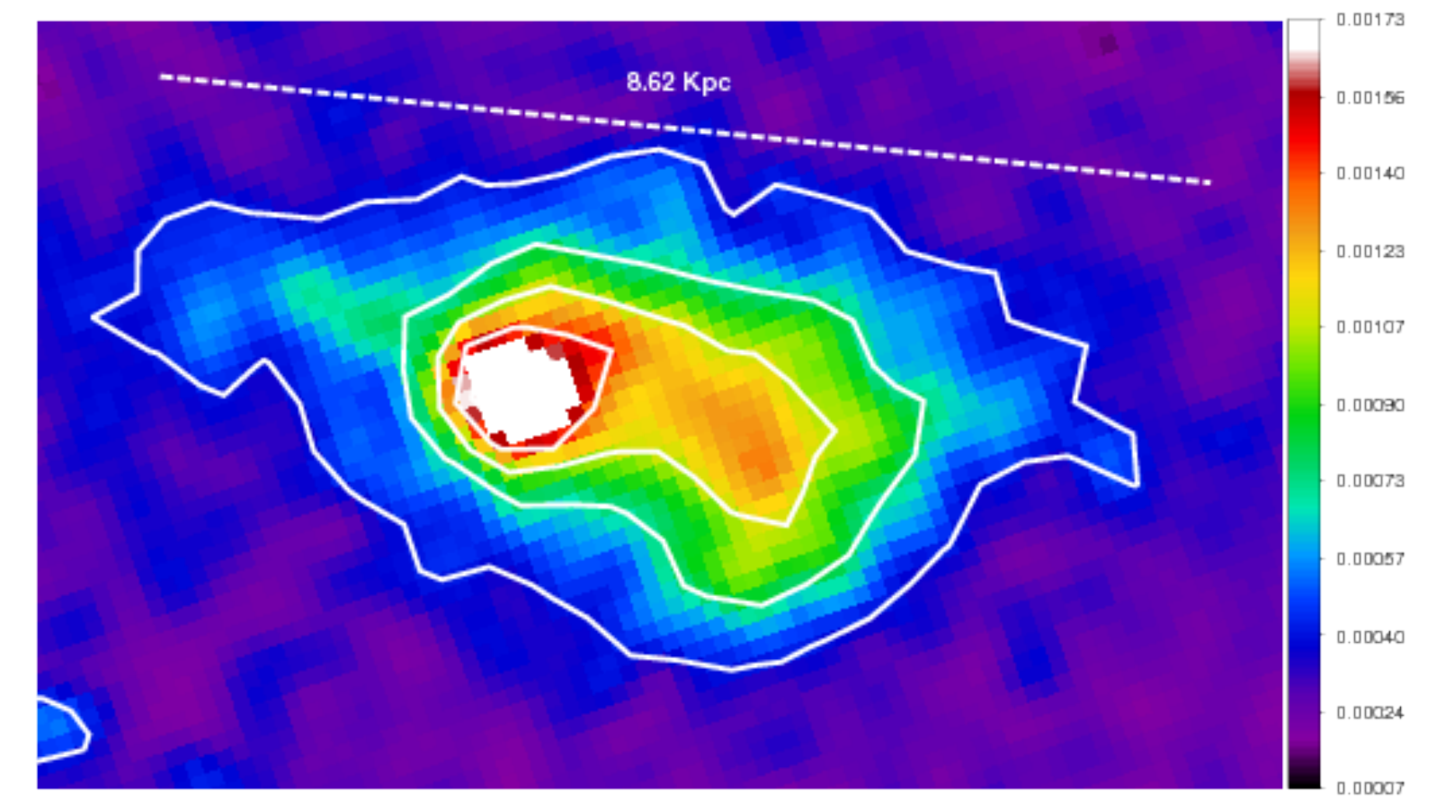}
  \caption{B}
\end{subfigure}
\begin{subfigure}{0.3\textwidth}
  \centering
  \includegraphics[width=1.\linewidth]{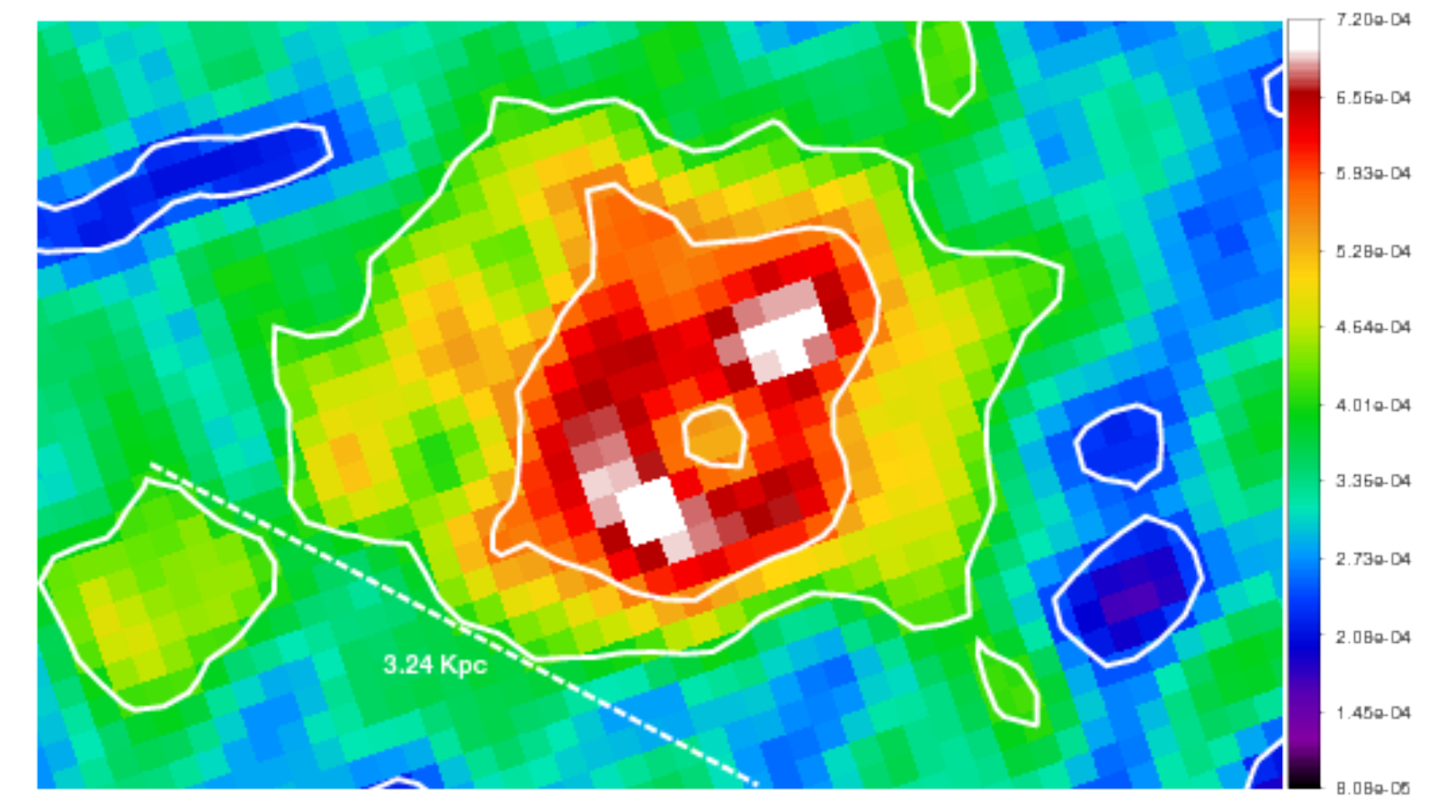}
  \caption{C}
\end{subfigure}
\begin{subfigure}{0.3\textwidth}
  \centering
  \includegraphics[width=1.\linewidth]{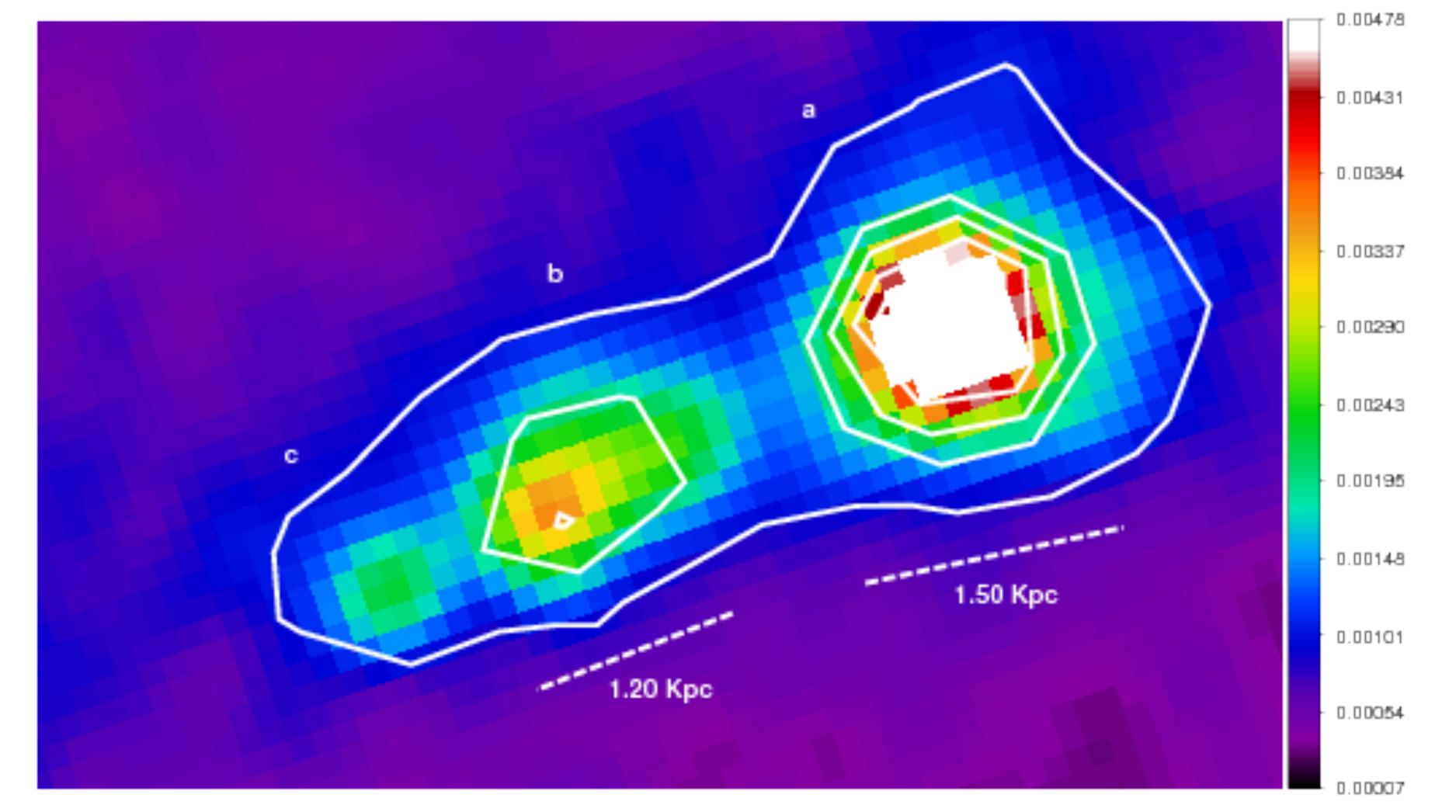}
  \caption{E}
\end{subfigure}
\begin{subfigure}{0.3\textwidth}
  \centering
  \includegraphics[width=1.\linewidth]{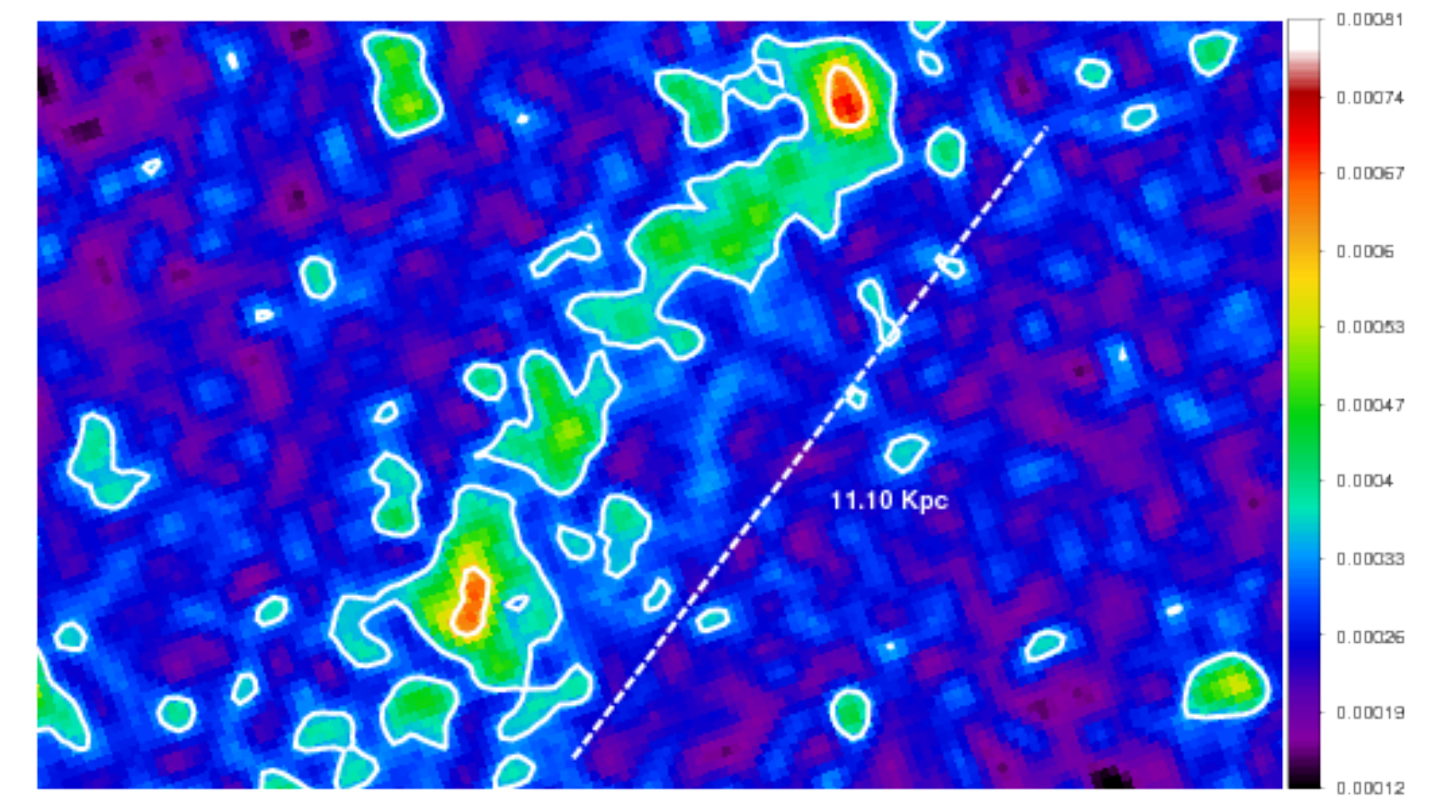}
  \caption{F}
\end{subfigure}%
\begin{subfigure}{0.3\textwidth}
  \centering
  \includegraphics[width=1.\linewidth]{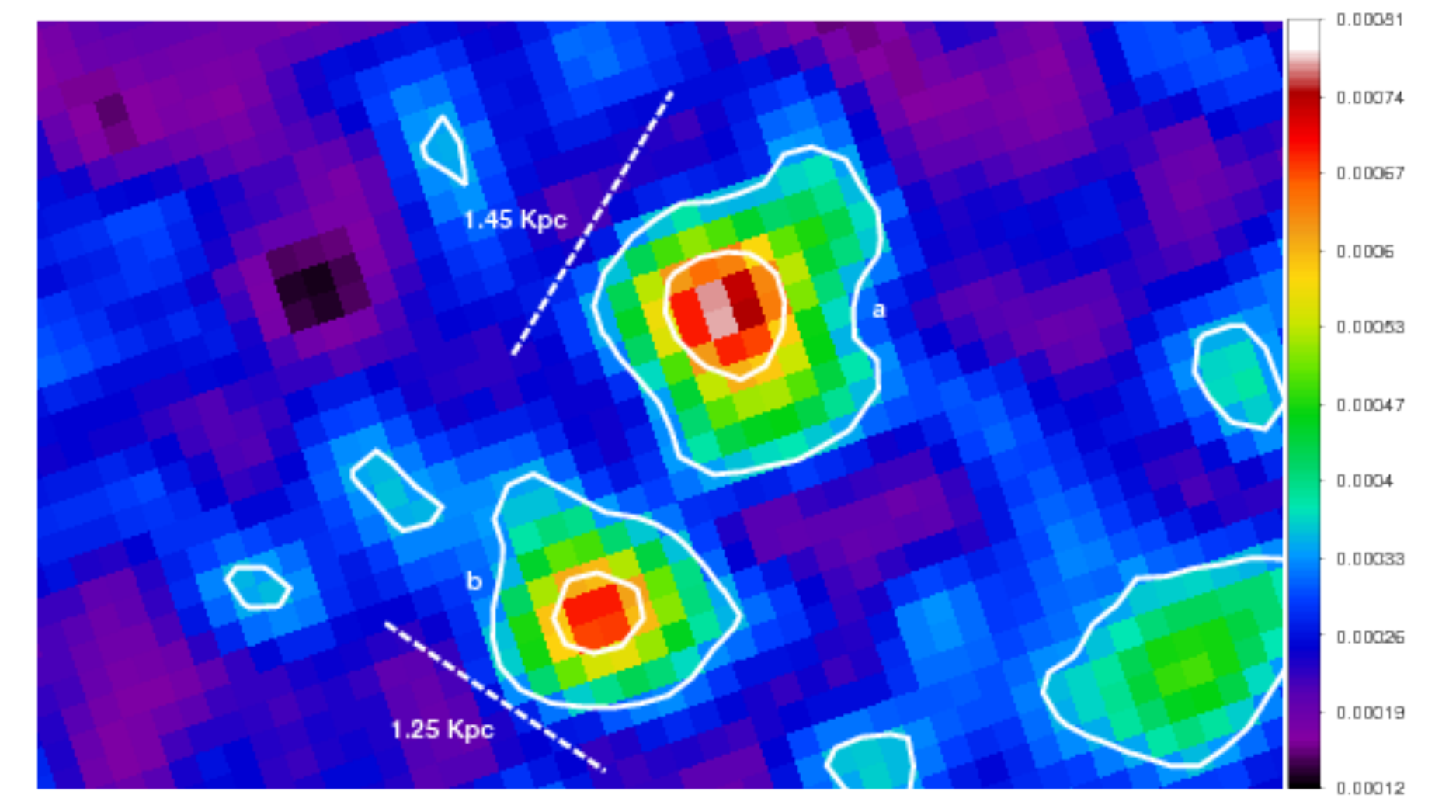}
  \caption{G}
\end{subfigure}
\begin{subfigure}{0.3\textwidth}
  \centering
  \includegraphics[width=1.\linewidth]{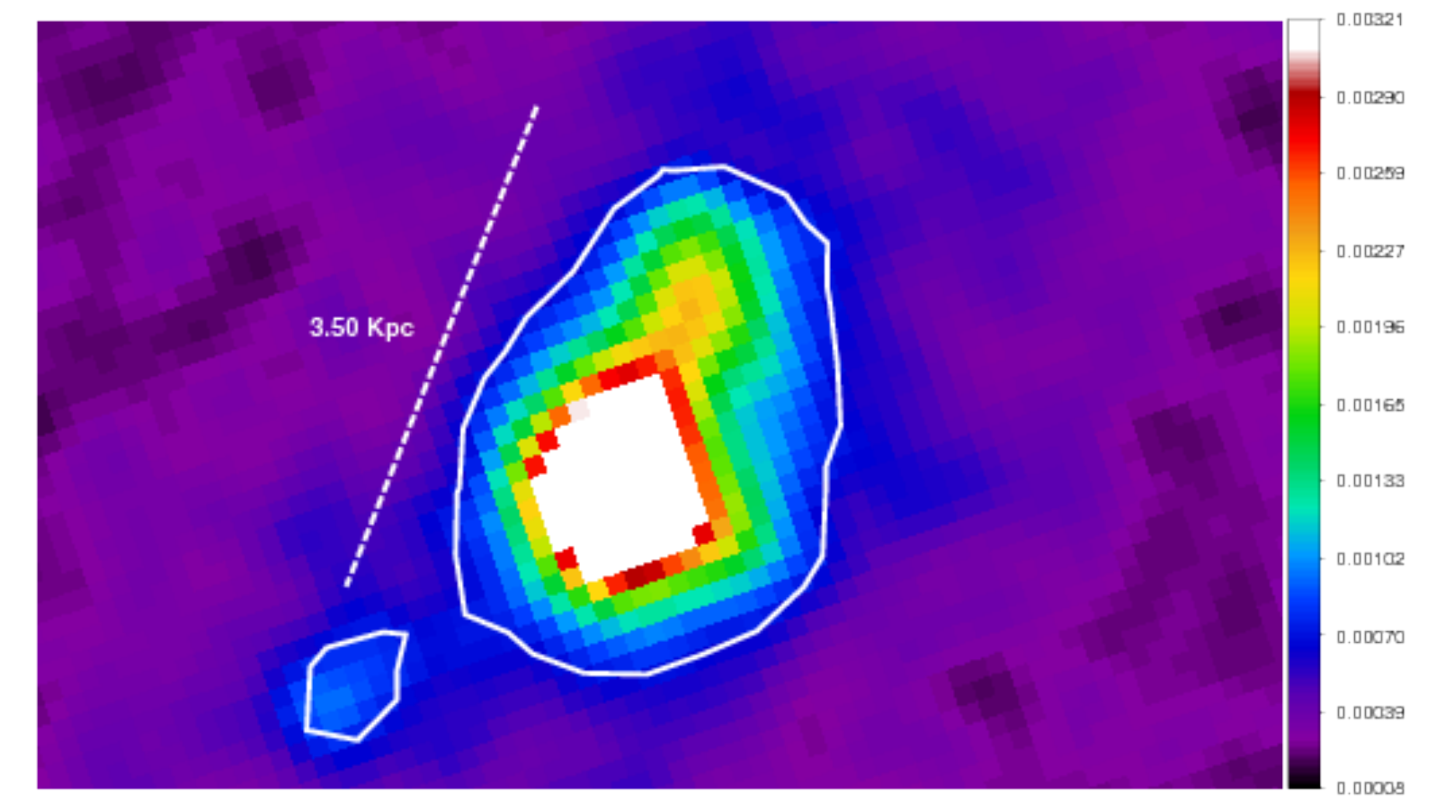}
  \caption{H}
\end{subfigure}


\caption{Star forming regions in the tidal tails of NGC7252 from the NUV image is shown in the boxes from sub fig. 1-to-7 (corresponding to A-to-H in Figure~\ref{figure:fig2}, excluding the central region (D) of galaxy). The scaling is chosen for each box to enhance the contrast to show the flux changes within the star forming regions. A colour bar is shown with scaling in counts per second.}
\label{figure:fig3}
\end{figure*}

The sites of ongoing and recent star formation in NGC7252 can be identified from the UV images, resolving features at 0.38 Kpc (1\farcs2) in the NUV and at 0.45  Kpc (1\farcs4) in the FUV.  The NUV spatial resolution is best suited to the identification of star forming regions in NGC7252 and we use it to mark the sites of ongoing star formation in the tidal tails of NGC7252. Figure~\ref{figure:fig2} displays the NUV image with the contrast level chosen to highlight  low surface brightness features in the tidal tails. The known tidal  features are marked and the colour scaling is shown in counts per second. It is to be noted that the size of few of the star forming regions in the tidal tails are similar to nearby dwarf galaxies. The known TDG  candidates (NGC7252 E and NGC7252 W) are located at the ends of the eastern and north western tidal tails.
The candidate star forming regions in tidal tails of NGC7252 are highlighted by boxes in Figure~\ref{figure:fig3}.  The scaling is changed for each box to enhance the contrast which, in turn, shows the NUV  flux changes within the star forming region. Contours are used to isolate obvious features.

We see the following details in Figure~\ref{figure:fig3}: (A) This can be a star forming region  associated with NGC7252. (B) The candidate tidal dwarf galaxy (NGC7252E) \citep{Lelli_2015}. The star forming region is elongated with intense NUV emission from the  eastern edge.  (C) The features  has intense NUV emission from a ring and a  different morphology in the optical $B-$band image. (E) The features consists of three  regions with different sizes. The features are just detached from the main body of the galaxy. (F) and (G) The features are candidate star forming  clumps which demonstrates small-scale star formation in regions along the tidal tails. (H) The feature  is a candidate tidal dwarf galaxy (NGC7252NW) \citep{Lelli_2015}. The intense NUV emission is seen  from the southern region of NGC7252NW. The north western tidal tail may have material falling back to the galaxy \citep{Hibbard_1995}. We note that NGC7252NW has a gradient in flux  towards the main body of galaxy: this could be due to star formation in the tidal material falling back towards the main galaxy.

\subsection{Star formation rates}
The FUV flux from the star forming regions in galaxies is a direct tracer of current star formation. The FUV flux can be used to compute the star formation rate of star forming regions assuming a constant star formation rate over the past 10$^8$ years. Recent simulations suggests that the derived star formation rate using estimators other than Lyman continuum can suffer from instantaneous changes over small ($\sim$ 10 Myr) timescales (see \citet{Boquien_2014}). The star formation rate along the tidal tails and the main body of NGC7252 is computed for a Salpeter initial mass function from the FUV luminosity (L$_{FUV}$) \citep{Kennicutt_1998}. We used the following form of equation as described in \citet{Iglesias_2006} and adopted in \citet{Cortese_2008}.

\begin{equation}
SFR_{FUV} [M_{\odot}/yr]  = \frac{L_{FUV}[erg/sec]} {3.83 \times 10^{33}} \times {10^{-9.51}} 
\end{equation}

\begin{figure}
\centering
\includegraphics[width=9cm,height=9cm,keepaspectratio]{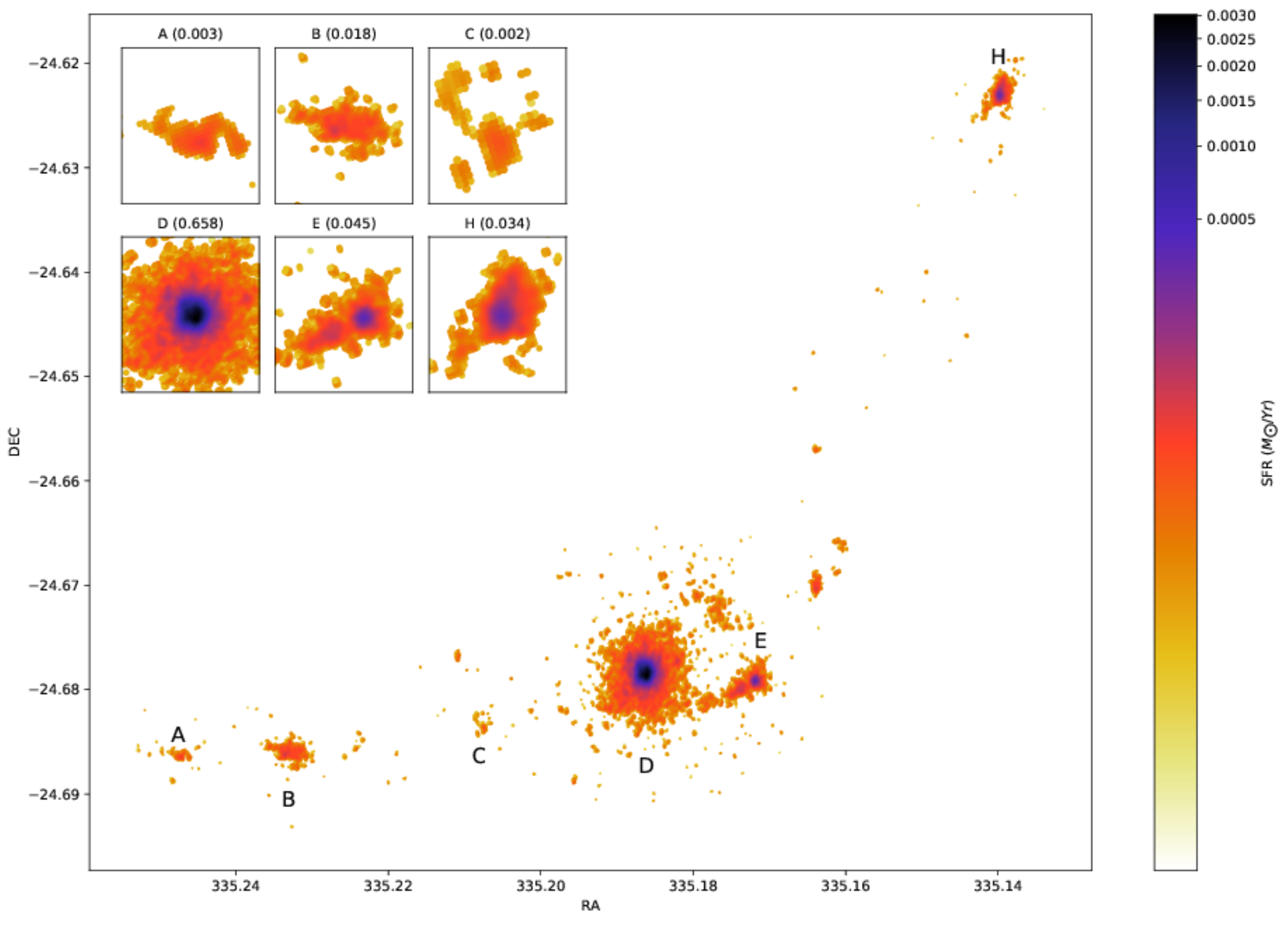}
\caption{The star formation rate for NGC7252. The spatial variation of star formation in tidal tails and the main body of galaxy is seen in the magnified regions within the boxes. The features are marked with  labels for identification and the star formation rate of each feature is noted at the top of the box.}\label{figure:fig4}
\end{figure}

The region of the FUV image that corresponds to the tidal tails and main body of NGC7252 was isolated using the background counts from the whole image to set the threshold. Pixels with values  above the 3$\sigma$ of the threshold were selected to isolate the region of interest. The counts in the selected pixels were background subtracted, integration time weighted and converted to flux units using the unit conversion (UC) (see Table 4 of \citep{Tandon_2017b} for the UC). The FUV flux of the selected region is converted to luminosity and is used to calculate the star formation rate. The spatial map of star formation rate along  the tidal tail and the galaxy is shown in Figure~\ref{figure:fig4}. The integrated star formation rate for NGC7252 in its entirety is found to be 0.81 $\pm$ 0.01 M$_{\odot}$/yr. Note that the integrated star formation rate  estimated from far infrared flux is 8.1 M$_{\odot}$/yr \citep{OSullivan_2015}. But see also Table 14 of \citet{Boquien_2009} where the integrated star formation rates derived from GALEX FUV flux of few selected regions in the tidal tail is found to be 0.08 $\pm$ 0.01. The extinction-free estimate of the SFR measured from 1.4GHz radio continuum observation of the central nucleus region of NGC7252 is found to be 6.3  $\pm$ 0.2 M$_{\odot}$/yr. The SFR from integrated $\mathrm{H}{\alpha}$ luminosity corrected for internal extinction (Av=1.75 $\pm$ 0.25 mag) is found to be 5.6  $\pm$ 1.1 M$_{\odot}$/yr \citep{Schweizer_2013}. The star forming regions thus measured, without any extinction correction, in the tidal tails and main body of the galaxy are shown in Figure~\ref{figure:fig4} and also magnified in boxes. Note that each point in Figure~\ref{figure:fig4} corresponds to 1  pixel in the FUV image and the integrated star formation rates are marked on the top of the boxes in the figure.

\begin{figure}
\centering
\includegraphics[width=9.5cm,height=9.5cm,keepaspectratio]{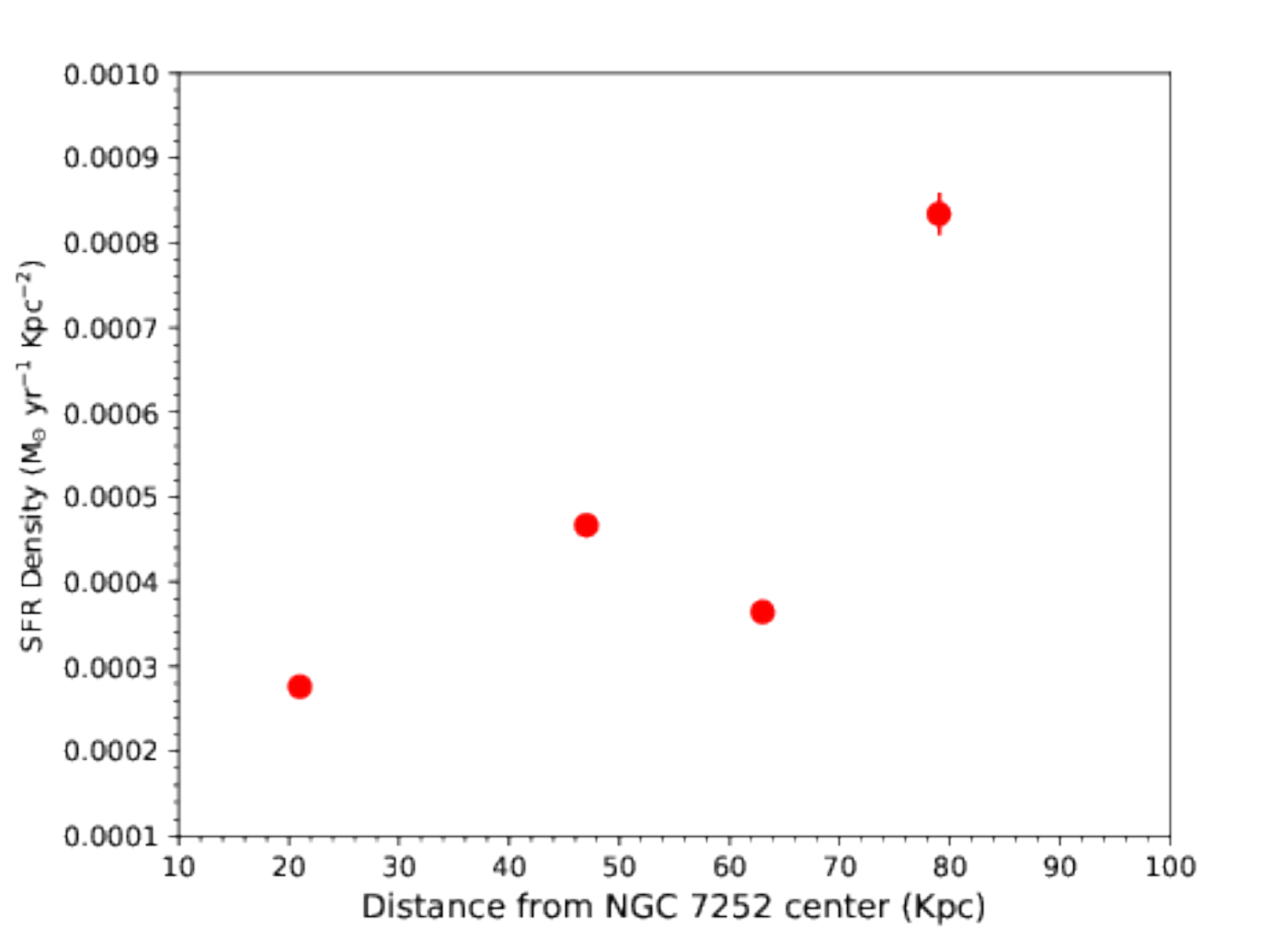}
\caption{The distance dependence on the integrated  star formation rate density (star formation rate per unit area) of star forming regions on the tidal tail from the centre of NGC7252 is shown.}\label {figure:fig5}
\end{figure}

The integrated star formation rate of the tidal features and area of the star forming regions  marked at  Figure~\ref{figure:fig4} is given in Table~\ref{Starformationrates}. The regions marked "B" and "H" in Figure~\ref{figure:fig4} are the candidate TDGs (NGC7252E, NGC7252NW) in NGC7252. The main body of the galaxy (the merger remnant) shows higher star formation rates (0.66 $\pm$ 0.01 M$_{\odot}$/yr) and shows spiral like features emanating from the centre region. This is clearly seen in optical from HST observations of the central region of NGC7252 \citep{Miller_1997,Laine_2003,Rossa_2007}. There will be FUV emission from any AGN at the centre of the merger remnant which would overestimate our star formation rate estimation \citep{Schweizer_2013}. The star  forming regions (marked B,F,H in Figure~\ref{figure:fig4}) show knots of intense star formation. If star formation arises in gas expelled during the merger, there can be a dependence of star formation rates  with distance from the main body of the galaxy. In  Figure~\ref{figure:fig5}, we confirm this variation in star formation rate surface density (star formation rate normalized to surface area) with distance from the centre of NGC7252. The most distant regions from this post-merger galaxy clearly show higher star formation rates.



\begin{table}
\caption{\label{t7} The star formation rate and area of the regions shown in Figure~\ref{figure:fig4}.}
\label{Starformationrates}
\tabcolsep=0.05cm
\begin{tabular}{ccl} 
\hline
ID & SFR (M$_{\odot}$/yr) & Area (Kpc$^{2}$)\\
\hline
A    & 0.003   & 8  \\
B    & 0.020  & 39  \\
C    &0.002  & 7   \\
D    &0.66   & 334 \\
E    &0.04   & 62  \\
H    &0.03   & 40  \\
\hline
\end{tabular}
\end{table}

\section{Discussion}

We have studied recent star formation in the post-merger galaxy NGC7252. The galaxy, a merger remnant, is understood to have formed from a major merger between two gas rich galaxies. Major and minor mergers of galaxies are known to trigger star formation even in non star-forming elliptical galaxies \citep{George_2015,George_2017}. Galaxy mergers are known to produce intense star formation episodes in the remnant as well as their tidal tails. Since star formation in normal galaxies happens primarily in their disks, such tidal tails provide an opportunity to study star formation in low-density environments. The tidal tails are composed of stellar populations with gas and dust content similar to the galaxy disk.

The star formation rate measured from the observed FUV flux for the main body of the galaxy is much smaller (SFR:0.66 M$_{\odot}$/yr) compared to the values obtained from Far-IR/1.4GHz/$\mathrm{H}{\alpha}$ luminosities (SFR:8.1/6.3/5.6 M$_{\odot}$/yr) \citep{Schweizer_2013}. The central star forming regions of NGC7252 can contain significant amount of dust that could attenuate the flux at UV wavelengths. The FUV flux we measured is not corrected for extinction and can be the reason for an $\sim$ order of magnitude discrepancy in SFR derived using FUV and other proxies. We try to compute the extinction in FUV (A$_{FUV}$) for the main body of the galaxy by considering the $\mathrm{H}{\alpha}$ derived SFR for the central region of the galaxy as the ideal one. We found that the corresponding value of A$_{FUV}$ = 2.33 mag. This implies that the observed FUV flux can be attenuated by a factor $\sim$ 8.5. If we consider a \citet{cardelli_1989} extinction law this corresponds to A$_{V}$=0.90 mag.

Multi-wavelength observations of the post-merger galaxy NGC7252 have been carried out from radio to X-rays in studies of its gas content, dynamics and star formation properties \citep{Schweizer_1982,Dupraz_1990,Borne_1991,Richter_1994,Hibbard_1995,Read_1998}. The present study has, for the first time, derived directly the star formation properties of this galaxy from wide-field UV images. The ongoing star-formation in the tidal tails of NGC7252 is directly observed  using the FUV/NUV images from UVIT. The features along tidal tails are hosting intense star formation with a rate density showing a mild dependence on distance from the central galaxy. Note that the observed trend is based on four star forming regions along the tidal tails and given the uncertainties from projection effects, should be taken with caution. The intense star formation in the tails can be due to gravitational perturbations or recent shocks in the gas ejected during the
merger \citep{Elmegreen_1996,Struck_1997,Smith_2008,Boquien_2009}. An obvious question is if intrinsically true, what drives the distance dependence on star formation density along the tidal tails? There can be different initial conditions responsible for the condensation of molecular gas that triggered star formation in the intergalactic medium. The gas along the tidal tails may have been expelled at different times during the merger process or, alternately, of clumps of differing mass. These initial conditions might have been imprinted on the star formation rates by the distance dependence of star formation surface density. Pre-merger disk galaxies with different velocity components are shown to later become the post-merger galaxy NGC7252 \citep{Hibbard_1995}. The gas expelled during the merger process can have the memory of the kinematics from the parent galaxies with different gas condensation time scales. The observed star formation rate density dependence with distance in the tidal tails of NGC7252 can then be explained by gas expelled at different times during the merging of two disk galaxies.\\

\begin{figure}
\centering
\includegraphics[width=8.5cm,height=8.5cm,keepaspectratio]{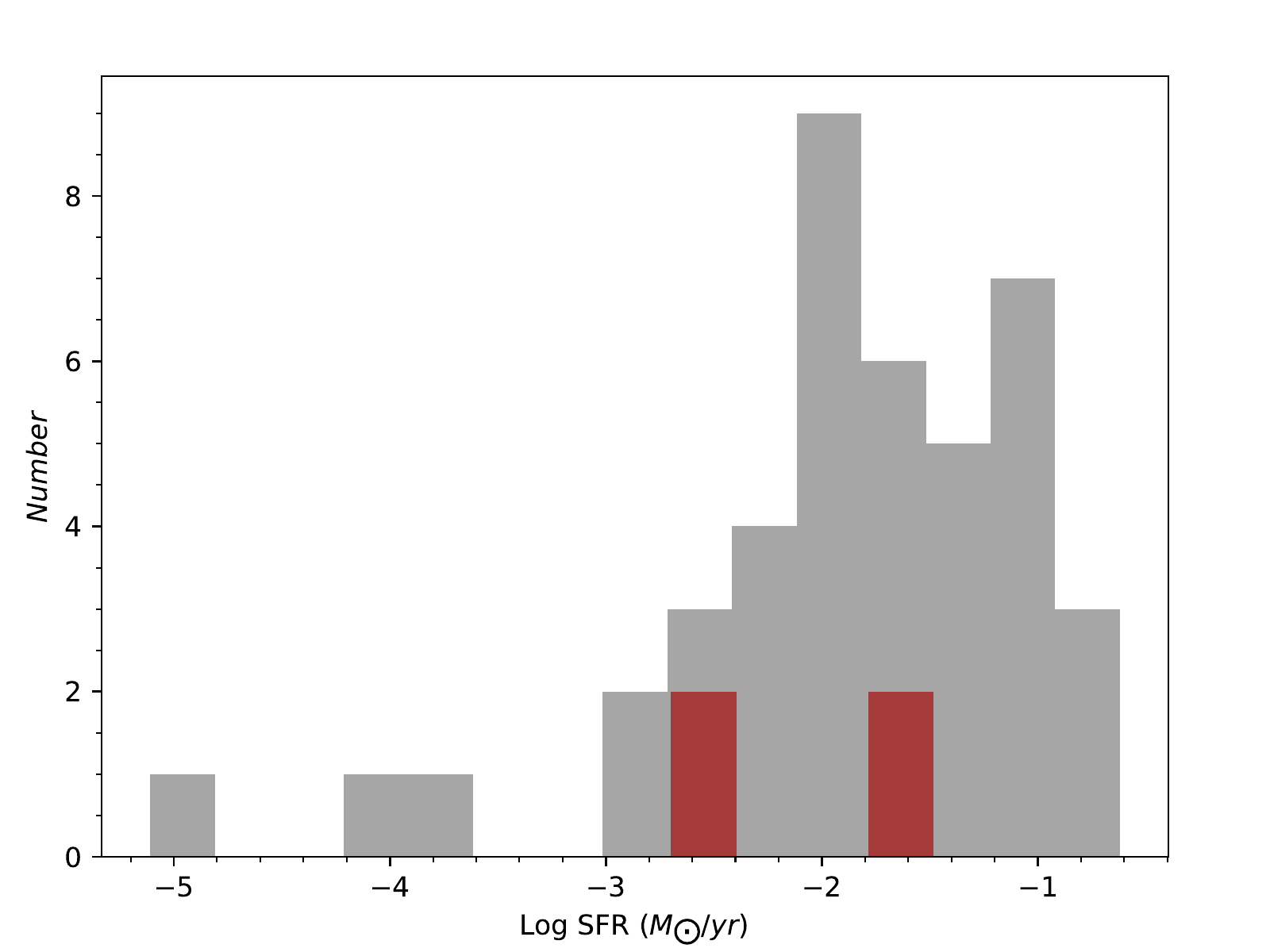}
\caption{The star formation rates of the star forming regions in the tidal tail of NGC7252 are shown in red. The star formation rate distribution for dwarf galaxies in the local Universe from \citet{Hunter_2010} are shown in grey.}\label{figure:fig7}
\end{figure}

The candidate TDGs found along the tidal tail of NGC7252 show intense star formation. Their star formation rates are comparable with those of typical dwarf galaxies in the local Universe. The star formation rate distribution of a sample of 44 dwarf galaxies (dwarf irregular, blue  compact  dwarf, and  Sm  galaxies) measured  from  integrated GALEX FUV flux is shown in Figure~\ref{figure:fig7} (in grey) along with the star formation rate of the star forming regions in the tidal tail (in red) (local Universe dwarf galaxy SFR from \citet{Hunter_2010}. One may question whether the TDGs are newly formed galaxies in the tidal tail or just dwarf galaxies that happen to be in the field. The TDGs on the tidal tail are observed to have solar gas phase metallicity (12 ${+}$ log (O/H) ${=}$ 8.6). This is a strong evidence towards the formation of dwarf galaxy from the pre-enriched gas on the tidal tails \citep{Lelli_2015}. We also note that there is spatial gradient in star formation rate within the TDGs. This can be due to the clumpy nature of triggered star formation happening due to gas condensation within the pre-enriched gas thrown out of the galaxies during the merger process. The progenitors of at least few (if not all) dwarf elliptical galaxies in the local Universe are proposed to be TDGs at higher redshifts \citep{Dabringhausen_2013}. \\

\section{Summary}
We have examined the spatially-resolved star formation properties of the post-merger galaxy NGC7252 using UV images acquired with the UVIT instrument onboard Astrosat. We identified recent star formation activity in NUV and FUV images, and derived star formation rates along the tidal tails and main body of NGC7252. The integrated star formation rate of NGC7252 computed with out extinction correction is found to be 0.81 $\pm$ 0.01 M$_{\odot}$/yr. The extinction in the central region of the galaxy assuming similar intrinsic SFR for the FUV and $\mathrm{H}{\alpha}$ proxies is found to be $\sim$ 2.33 $mag$. Significant star formation is detected in the tidal dwarf galaxy candidates (NGC7252E, SFR=0.02   M$_{\odot}$/yr and NGC7252NW, SFR=0.03 M$_{\odot}$/yr) which is comparable to the integrated star formation rates observed in dwarf galaxies in local Universe. The star formation rate density of star forming regions in tidal tails seem to show a dependence on distance from the galaxy centre. This trend may be due to the difference in initial conditions responsible for the triggering of star formation in the molecular gas that was expelled during the recent merger in NGC7252.



 
 

\begin{acknowledgements}
We thank the referee for his/her comments that improved the scientific content of the paper. KG thank Smitha Subramanian for carefully reading the manuscript and providing comments. This publication uses the data from the AstroSat mission of the Indian Space Research  Organisation  (ISRO),  archived  at  the  Indian  Space  Science  Data 
Centre (ISSDC).  UVIT  project  is  a  result  of collaboration  between  IIA,  Bengaluru,  IUCAA,  Pune, TIFR, Mumbai, several centres of ISRO, and CSA. Indian Institutions and the Canadian Space Agency have contributed to the work presented in this paper.  Several  groups  from  ISAC  (ISRO),  Bengaluru,  and  IISU (ISRO), Trivandrum have contributed to the design, fabrication, and testing of the payload.  The Mission Group (ISAC), ISTRAC (ISAC), UVIT-POC (IIA) continue to provide support in making observations with, and reception and initial processing of the data.  We gratefully thank all the individuals involved in the various teams for providing their support to the project from the early stages of the design to launch and observations with it in the orbit.
\end{acknowledgements}



\begin{thebibliography}{}

\bibitem[Agrawal(2006)]{Agrawal_2006} Agrawal, P.~C.\ 2006, Advances in Space Research, 38, 2989 

\bibitem[Baldry et al.(2004)]{Baldry_2004} Baldry, I.~K., Glazebrook, K., Brinkmann, J., et al.\ 2004, \apj, 600, 681 

\bibitem[Barnes \& Hernquist(1992)]{Barnes_1992} Barnes, J.~E., \& Hernquist, L.\ 1992, \nat, 360, 715 

\bibitem[Bastian et al.(2013)]{Bastian_2013} Bastian, N., Schweizer, F., Goudfrooij, P., Larsen, S.~S., \& Kissler-Patig, M.\ 2013, \mnras, 431, 1252 

\bibitem[Bell et al.(2004)]{Bell_2004} Bell, E.~F., Wolf, C., Meisenheimer, K., et al.\ 2004, \apj, 608, 752 


\bibitem[Blumenthal et al.(1984)]{Blumenthal_1984} Blumenthal, G.~R., Faber, S.~M., Primack, J.~R., \& Rees, M.~J.\ 1984, \nat, 311, 517 

\bibitem[Boquien et al.(2007)]{Boquien_2007} Boquien, M., Duc, P.-A., Braine, J., et al.\ 2007, \aap, 467, 93 

\bibitem[Boquien et al.(2009)]{Boquien_2009} Boquien, M., Duc, P.-A., Wu, Y., et al.\ 2009, \aj, 137, 4561 

\bibitem[Boquien et al.(2014)]{Boquien_2014} Boquien, M., Buat, V., \& Perret, V.\ 2014, \aap, 571, A72 

\bibitem[Borne \& Richstone(1991)]{Borne_1991} Borne, K.~D., \& Richstone, D.~O.\ 1991, \apj, 369, 111 

\bibitem[Braine et al.(2000)]{Braine_2000} Braine, J., Lisenfeld, U., Due, P.-A., \& Leon, S.\ 2000, \nat, 403, 867 

\bibitem[Braine et al.(2001)]{Braine_2001} Braine, J., Duc, P.-A., Lisenfeld, U., et al.\ 2001, \aap, 378, 51 

\bibitem[Cardelli et al.(1989)]{cardelli_1989} Cardelli, J.~A., Clayton, G.~C., \& Mathis, J.~S.\ 1989, \apj, 345, 245 


\bibitem[Chien \& Barnes(2010)]{Chien_2010} Chien, L.-H., \& Barnes, J.~E.\ 2010, \mnras, 407, 43 

\bibitem[Cortese et al.(2008)]{Cortese_2008} Cortese, L., Gavazzi, G., \& Boselli, A.\ 2008, \mnras, 390, 1282 




\bibitem[Dabringhausen \& Kroupa(2013)]{Dabringhausen_2013} Dabringhausen, J., \& Kroupa, P.\ 2013, \mnras, 429, 1858 

\bibitem[Dopita et al.(2002)]{Dopita_2002} Dopita, M.~A., Pereira, M., Kewley, L.~J., \& Capaccioli, M.\ 2002, \apjs, 143, 47 


\bibitem[Duc \& Mirabel(1994)]{Duc_1994} Duc, P.-A., \& Mirabel, I.~F.\ 1994, \aap, 289, 83 


\bibitem[Duc \& Mirabel(1998)]{Duc_1998} Duc, P.-A., \& Mirabel, I.~F.\ 1998, \aap, 333, 813 

\bibitem[Duc et al.(2000)]{Duc_2000} Duc, P.-A., Brinks, E., Springel, V., et al.\ 2000, \aj, 120, 1238 

\bibitem[Duc et al.(2007)]{Duc_2007} Duc, P.-A., Braine, J., Lisenfeld, U., Brinks, E., \& Boquien, M.\ 2007, \aap, 475, 187 

\bibitem[Duc(2012)]{Duc_2012} Duc, P.-A.\ 2012, Astrophysics and Space Science Proceedings, 28, 305 

\bibitem[Dupraz et al.(1990)]{Dupraz_1990} Dupraz, C., Casoli, F., Combes, F., \& Kazes, I.\ 1990, \aap, 228, L5 

\bibitem[Elmegreen \& Efremov(1996)]{Elmegreen_1996} Elmegreen, B.~G., \& Efremov, Y.~N.\ 1996, \apj, 466, 802 



\bibitem[Faber et al.(2007)]{Faber_2007} Faber, S.~M., Willmer, C.~N.~A., Wolf, C., et al.\ 2007, \apj, 665, 265 



\bibitem[Fritze-v.~Alvensleben \& Gerhard(1994)]{Fritze_1994} Fritze-v.~Alvensleben, U., \& Gerhard, O.~E.\ 1994, \aap, 285, 751 





\bibitem[Genzel et al.(2001)]{Genzel_2001} Genzel, R., Tacconi, L.~J., Rigopoulou, D., Lutz, D., \& Tecza, M.\ 2001, \apj, 563, 527 

\bibitem[George \& Zingade(2015)]{George_2015} George, K., \& Zingade, K.\ 2015, \aap, 583, A103 


\bibitem[George(2017)]{George_2017} George, K.\ 2017, \aap, 598, A45 



\bibitem[Gil de Paz et al.(2007)]{Gildepaz_2007} Gil de Paz, A., Boissier, S., Madore, B.~F., et al.\ 2007, \apjs, 173, 185 

\bibitem[Girish et al.(2017)]{Girish_2017} Girish, V., Tandon, S.~N., Sriram, S., Kumar, A., \& Postma, J.\ 2017, Experimental Astronomy, 43, 59 

\bibitem[Hibbard et al.(1994)]{Hibbard_1994} Hibbard, J.~E., Guhathakurta, P., van Gorkom, J.~H., \& Schweizer, F.\ 1994, \aj, 107, 67 

\bibitem[Hibbard \& Mihos(1995)]{Hibbard_1995} Hibbard, J.~E., \& Mihos, J.~C.\ 1995, \aj, 110, 140 

\bibitem[Hibbard \& Yun(1999)]{Hibbard_1999} Hibbard, J.~E., \& Yun, M.~S.\ 1999, \apjl, 522, L93 

\bibitem[Hunter et al.(2010)]{Hunter_2010} Hunter, D.~A., Elmegreen, B.~G., \& Ludka, B.~C.\ 2010, \aj, 139, 447 

\bibitem[Iglesias-P{\'a}ramo et al.(2006)]{Iglesias_2006} Iglesias-P{\'a}ramo, J., Buat, V., Takeuchi, T.~T., et al.\ 2006, \apjs, 164, 38 


\bibitem[Kaviraj et al.(2012)]{Kaviraj_2012} Kaviraj, S., Darg, D., Lintott, C., Schawinski, K., \& Silk, J.\ 2012, \mnras, 419, 70 


\bibitem[Kennicutt(1998)]{Kennicutt_1998} Kennicutt, R.~C., Jr.\ 1998, \apj, 498, 541 

\bibitem[Kennicutt \& Evans(2012)]{Kennicutt_2012} Kennicutt, R.~C., \& Evans, N.~J.\ 2012, \araa, 50, 531 

\bibitem[Knierman et al.(2003)]{Knierman_2003} Knierman, K.~A., Gallagher, S.~C., Charlton, J.~C., et al.\ 2003, \aj, 126, 1227 

\bibitem[{{Komatsu} {et~al}\mbox{.}(2011){Komatsu}, {Smith}, {Dunkley},
  {Bennett}, {Gold}, {Hinshaw}, {Jarosik}, {Larson}, {Nolta}, {Page},
  {Spergel}, {Halpern}, {Hill}, {Kogut}, {Limon}, {Meyer}, {Odegard}, {Tucker},
  {Weiland}, {Wollack}, \& {Wright}}]{Komatsu_2011}
{Komatsu} E. {et~al.}, 2011, \apjs, 192, 18


\bibitem[Kroupa et al.(2010)]{Kroupa_2010} Kroupa, P., Famaey, B., de Boer, K.~S., et al.\ 2010, \aap, 523, A32 

\bibitem[Kumar et al.(2012)]{Kumar_2012} Kumar, A., Ghosh, S.~K., Hutchings, J., et al.\ 2012, \procspie, 8443, 84431N 


\bibitem[Lake \& Dressler(1986)]{Lake_1986} Lake, G., \& Dressler, A.\ 1986, \apj, 310, 605 

\bibitem[Laine et al.(2003)]{Laine_2003} Laine, S., van der Marel, R.~P., Rossa, J., et al.\ 2003, \aj, 126, 2717 


\bibitem[Lang et al.(2010)]{Lang_2010} Lang, D., Hogg, D.~W., Mierle, K., Blanton, M., \& Roweis, S.\ 2010, \aj, 139, 1782 


\bibitem[Lelli et al.(2015)]{Lelli_2015} Lelli, F., Duc, P.-A., Brinks, E., et al.\ 2015, \aap, 584, A113 


\bibitem[Metz \& Kroupa(2007)]{Metz_2007} Metz, M., \& Kroupa, P.\ 2007, \mnras, 376, 387 

\bibitem[Miller et al.(1997)]{Miller_1997} Miller, B.~W., Whitmore, B.~C., Schweizer, F., \& Fall, S.~M.\ 1997, \aj, 114, 2381 

\bibitem[O'Sullivan et al.(2015)]{OSullivan_2015} O'Sullivan, E., Combes, F., Hamer, S., et al.\ 2015, \aap, 573, A111 

\bibitem[Postma \& Leahy(2017)]{Postma_2017} Postma, J.~E., \& Leahy, D.\ 2017, \pasp, 129, 115002 


\bibitem[Read \& Ponman(1998)]{Read_1998} Read, A.~M., \& Ponman, T.~J.\ 1998, \mnras, 297, 143 

\bibitem[Richter et al.(1994)]{Richter_1994} Richter, O.-G., Sackett, P.~D., \& Sparke, L.~S.\ 1994, \aj, 107, 99 

\bibitem[Rossa et al.(2007)]{Rossa_2007} Rossa, J., Laine, S., van der Marel, R.~P., et al.\ 2007, \aj, 134, 2124 

\bibitem[Rothberg \& Joseph(2006)]{Rothberg_2006} Rothberg, B., \& Joseph, R.~D.\ 2006, \aj, 131, 185 


\bibitem[Schlegel et al.(1998)]{Schlegel_1998} Schlegel, D.~J., Finkbeiner, D.~P., \& Davis, M.\ 1998, \apj, 500, 525 


\bibitem[Schweizer(1978)]{Schweizer_1978} Schweizer, F.\ 1978, Structure and Properties of Nearby Galaxies, 77, 279 

\bibitem[Schweizer(1982)]{Schweizer_1982} Schweizer, F.\ 1982, \apj, 252, 455 

\bibitem[Schweizer et al.(2013)]{Schweizer_2013} Schweizer, F., Seitzer, P., Kelson, D.~D., Villanueva, E.~V., \& Walth, G.~L.\ 2013, \apj, 773, 148 


\bibitem[Smith et al.(2008)]{Smith_2008} Smith, B.~J., Struck, C., Hancock, M., et al.\ 2008, \aj, 135, 2406 


\bibitem[Struck(1997)]{Struck_1997} Struck, C.\ 1997, \apjs, 113, 269 

\bibitem[Subramaniam et al.(2016)]{Annapurni_2016} Subramaniam, A., Tandon, S.~N., Hutchings, J., et al.\ 2016, \procspie, 9905, 99051F 




\bibitem[Tandon et al.(2017)]{Tandon_2017a} Tandon, S.~N., Hutchings, J.~B., Ghosh, S.~K., et al.\ 2017, Journal of Astrophysics and Astronomy, 38, 28 

\bibitem[Tandon et al.(2017)]{Tandon_2017b} Tandon, S.~N., Subramaniam, A., Girish, V., et al.\ 2017, \aj, 154, 128 


\bibitem[Toomre \& Toomre(1972)]{Toomre_1972} Toomre, A., \& Toomre, J.\ 1972, \apj, 178, 623 

\bibitem[Toomre(1977)]{Toomre_1977} Toomre, A.\ 1977, Evolution of Galaxies and Stellar Populations, 401 

\bibitem[Visvanathan \& Sandage(1977)]{Visvanathan_1977} Visvanathan, N., \& Sandage, A.\ 1977, \apj, 216, 214

\bibitem[Wang et al.(1992)]{Wang_1992} Wang, Z., Schweizer, F., \& Scoville, N.~Z.\ 1992, \apj, 396, 510 

\bibitem[Whitmore et al.(1997)]{Whitmore_1997} Whitmore, B.~C., Miller, B.~W., Schweizer, F., \& Fall, S.~M.\ 1997, \aj, 114, 1797 

\bibitem[Zwicky(1956)]{Zwicky_1956} Zwicky, F.\ 1956, Ergebnisse der exakten Naturwissenschaften, 29, 344 

\end{thebibliography}
\end{document}